\documentclass[%
 reprint,
 amsmath,amssymb,
 aps,
]{revtex4-2}

\usepackage{graphicx}
\usepackage{amsmath}
\usepackage{lineno}
\usepackage{dcolumn}
\usepackage{bm}

\begin{document}

\preprint{APS/123-QED}

\title{Simple reactor model of relativistic runaway electron avalanche development}

\author{Egor Stadnichuk}
 \altaffiliation[1]{Moscow Institute of Physics and Technology - 1 “A” Kerchenskaya st., Moscow, 117303, Russian Federation}
 \altaffiliation[2]{HSE University - 20 Myasnitskaya ulitsa, Moscow 101000 Russia}
 \email{yegor.stadnichuk@phystech.edu}
 \affiliation{Moscow Institute of Physics and Technology - 1 “A” Kerchenskaya st., Moscow, 117303, Russian Federation\\
 HSE University - 20 Myasnitskaya ulitsa, Moscow 101000 Russia}
\author{Daria Zemlianskaya}%
 \email{zemlianskay.d@phystech.edu}
 \altaffiliation{%
 Moscow Institute of Physics and Technology - 1 “A” Kerchenskaya st., Moscow, 117303, Russian Federation\\
 Institute for Nuclear Research of RAS - prospekt 60-letiya Oktyabrya 7a, Moscow 117312
}%


\author{Ekaterina Svechnikova}
\email{svechnikova@ipfran.ru}
\altaffiliation{
 Institute of Applied Physics of RAS - 46 Ul'yanov str., 603950, Nizhny Novgorod, Russia
}%
\author{Eduard Kim}
 \altaffiliation{Moscow Institute of Physics and Technology - 1 “A” Kerchenskaya st., Moscow, 117303, Russian Federation\\ Institute for Nuclear Research of RAS - prospekt 60-letiya Oktyabrya 7a, Moscow 117312}
 \email{kim.e@phystech.edu}
 \author{Alexander Sedelnikov}
 \altaffiliation{Moscow Institute of Physics and Technology - 1 “A” Kerchenskaya st., Moscow, 117303, Russian Federation\\ Lebedev Physical Institute RAS}
\email{sedelnikov.as@phystech.edu}
\author{Oraz Anuaruly}
 \altaffiliation{Moscow Institute of Physics and Technology - 1 “A” Kerchenskaya st., Moscow, 117303, Russian Federation\\ Kurchatov Institute, Russian
Research Centre - sq. Academician Kurchatov, 1, Moscow, 123098, Russian Federation}
 \email{orazanuaruly@gmail.com}

\date{\today}

\begin{abstract}

High-energy gamma radiation in the Earth's atmosphere is associated with the bremsstrahlung of Relativistic Runaway Electron Avalanches (RREA) developing in thunderstorm electric fields. In this paper, RREA development is studied in the system of two strong electric-field regions within thunderstorms, which accelerate runaway electrons toward each other. Such a system is called the simple reactor. It is discovered that the propagation of gamma rays and runaway electrons from one region to another leads to positive feedback. This feedback called the reactor feedback can make RREA self-sustaining, thus effectively multiplying high-energy particles inside thunderstorms containing the simple reactor. The spectrum and characteristic time scale of the simple reactor gamma radiation are in agreement with Terrestrial Gamma-ray Flashes (TGFs) data. The applicability of the simple reactor model to TGF is discussed, and the distinguishing observable properties of the simple reactor radiation during TGF and Thunderstorm Ground Enhancement are considered.

\end{abstract}

\maketitle


\section{Keypoints}
\begin{itemize}
\item RREA development in thunderstorms containing the simple reactor is studied
\item Two feedback mechanisms that can make RREA self-sustaining in the simple reactor are discovered: electron and gamma-ray reactor feedback
\item Characteristics of gamma radiation of the simple reactor are in agreement with TGF and TGE experimental data
\end{itemize}

\section{\label{sec:level1}Introduction}

Atmospheric physics is rich in mysterious natural phenomena. One of the new directions in atmospheric research is high-energy atmospheric physics. It suddenly appeared in 1992, when the Burst and Transient Source Experiment (BATSE) detector aboard the Compton Gamma-Ray Observatory experiment discovered short and intensive bursts of gamma rays originating in the atmosphere of Earth \cite{BATSE_TGF_discovery}. These energetic bursts are called Terrestrial Gamma-ray Flashes (TGFs). It is established that the source of TGFs are thunderstorms \cite{Fermi_2016}. The characteristic duration of a TGF is 100~$\mu$s \cite{10_month_ASIM}, energy of detected TGF gamma-rays is up to 40~MeV \cite{asim_spectral_analysis,Fermi_first_results}. Thunderstorm gamma radiation is also detected on the Earth's surface. It is called Thunderstorm Ground Enhancement (TGE) \cite{Chilingarian_discovery} or gamma-ray glows \cite{Wada2019}, and its characteristic duration is up to tens of minutes. It is important to note that high-energy processes within thunderstorms are closely related to lightning. TGE precede lightning and are always terminated by lightning discharges \cite{Chilingarian_2020_radon}. TGFs are established to occur at the early stage of the lightning leader propagation \cite{asim_science,asim_tgf_lightning,Mezentsev_2022_TGF_lightning}. Moreover, many other interesting bright phenomena were registered in connection with the high-energy radiation from thunderstorms \cite{Enoto2017_nuclear_triggered,asim_tgf_elves,ASIM_first_TEB}.

The underlying physics of high-energy atmospheric radiation is the acceleration of electrons in thunderstorm electric fields \cite{Chilingarian_2011_natural_accelerator,Fermi_2016,Babich_2020,Dwyer2012_phenomena,Koehn_2017,TGF_particle_accelerator_2011}. In strong thunderstorm electric fields, relativistic electrons obtain more energy from the acceleration in the electric field than they on average lose on interactions with atmosphere air molecules. Such electrons are called runaway electrons \cite{Gurevich1992}. When the electric field strength exceeds the critical value, $E_c = 276 \frac{kV}{m \cdot atm}$, runaway electrons are Møller scattered by air molecules, which leads to the appearance of additional runaway electrons \cite{Dwyer_2003_fundamental_limit}. In this way, runaway electrons multiply in the process of their propagation along the thunderstorm electric field, forming the Relativistic Runaway Electron Avalanche (RREA) \cite{Babich_2020, Dwyer2012_phenomena}. To start a RREA an initial seed energetic particle is needed to appear within the thunderstorm electric field \cite{Babich_2020,CHILINGARIAN2021_particle_accelerator}. For example, it can be a secondary cosmic ray particle \cite{Gurevich_2001} or a seed particle generated inside the thunderstorm \cite{Moss_2006,Dwyer_vs_Gurevich,reactor}. Characteristic RREA particle energies range from tens of keV to tens of MeV \cite{Babich_2020}. Thus, runaway electrons naturally produce bremsstrahlung gamma rays in collisions with air molecules, which is detected as TGF and TGE \cite{Dwyer2012_phenomena,asim_spectral_analysis,asim_tgf_lightning,Chilingarian_2020_radon,CHILINGARIAN2021_particle_accelerator,Wada2019}.

The mystery of TGFs is that a large number of high-energy particles appear almost instantly inside a thundercloud \cite{BATSE_TGF_discovery,Fermi_first_results,Dwyer_vs_Gurevich}. There are two possible ways to explain such phenomena. The first possible scenario is the generation of a large number of seed electrons within the thunderstorm super strong electric fields, possibly created by the lightning leader propagation \cite{Koehn_2017,Moss_2006,Dwyer_vs_Gurevich,Kohn_2020_coronae,Kohn_2020_complex_streamer,Celestin2012}. Also it is considered that lightning leader itself can radiate synchrotron gamma-rays \cite{Petrov2021}. These ideas are supported by the fact that x-rays are observed in association with lightning leader propagation \cite{Dwyer_2005_xray_leader_step,Dwyer_2011_dart_leader_x_ray}. The second possible scenario is the multiplication of RREAs by positive feedback mechanisms \cite{Dwyer_2003_fundamental_limit,Dwyer2007,STADNICHUK_positron_criterion,Kutsyk2011606,reactor,Zelenyi_Dwyer,Skeltved2014}. The relativistic feedback works in the following way. Bremsstrahlung gamma ray radiated by runaway electrons can produce electron-positron pairs within thunderstorm supercritical electric field region. Positrons are accelerated by the electric field in the direction opposite to the runaway electrons acceleration direction. In this way, positrons reach the beginning of the supercritical region, where they produce seed runaway electrons by the Bhabha scattering \cite{Dwyer2007}. Thus, relativistic positron feedback multiplies RREAs and, moreover, can make RREA self-sustaining \cite{Dwyer_2003_fundamental_limit}. Similarly bremsstrahlung gamma rays can Compton backscatter and thus produce seed runaway electrons at the beginning of the supercritical region, which is the relativistic gamma ray feedback \cite{Dwyer_2012}. RREA models based on the positive feedback are supported by the fact that their characteristic time and spectrum coincide with the characteristic time and spectrum of TGF \cite{Dwyer_vs_Gurevich,Fermi_2016,asim_spectral_analysis,10_month_ASIM}. Nevertheless, the relativistic feedback requires strong large-scale electric fields, which have never been directly experimentally observed in thunderstorms \cite{STADNICHUK_positron_criterion,Stadnichuk2019,Zelenyi_Dwyer,Dwyer2007}.

It has been discovered that non-uniform thunderstorm electric field geometry leads to another feedback mechanism called the reactor feedback \cite{reactor,Zelenyi_reactor}. Let a thunderstorm consist of several separate electric-field regions with electric-field strength sufficient for RREA production. Such regions, for simplicity, are called cells \cite{reactor}. If a seed electron starts a RREA within one of the cells, the following processes occur. A RREA radiates bremsstrahlung gamma-rays. Gamma-rays have a large attenuation length at thunderstorm altitudes. Therefore, gamma-ray photons propagate through the thunderstorm and can reach other cells. There is a probability that a gamma-ray photon will interact with air molecules by compton scattering, photoelectric effect of electron-positron pair production within a cell, which can result in runaway electron generation. A runaway electron can produce a RREA. In this way the reactor gamma-ray feedback works: separate thunderstorm cells irradiate each other with gamma radiation, which results in RREA multiplication. Another reactor feedback mechanism - runaway electron transport between cells. If the cells are close to each other, runaway electrons are able to penetrate the air layer between them. In this way, runaway electrons propagate from one cell to another and, thus, multiply RREA. In general, reactor feedback is defined as the multiplication of RREA by high-energy particle exchange between separate thunderstorm RREA-accelerating regions. Distant cells amplify each other mostly by gamma-ray photon exchange because of their high penetrating power in the air. For cells located close to each other, the reactor feedback works mainly by runaway electron exchange, since RREA consists mainly of runaway electrons. It has been established that reactor feedback can lead to self-sustaining development of RREA, and, moreover, requires lower electric field strength in comparison with the relativistic feedback \cite{reactor}.

In this paper, the reactor feedback is studied in the simplest case of non-uniform thunderstorm electric field, when thunderstorm consists of two cells, oriented in the way that they accelerate runaway electrons towards each other. The system is called the simple reactor. The research is motivated by observations of thunderstorm electric structures, part of which can be described as the simple reactor electric structure \cite{Stolzenburg_1994, Marshall_1995, Marshall_1998_estimates, Stolzenburg_1998_structure_2, Stolzenburg_1998_structure_3, Stolzenburg_1998_precipitation, Stolzenburg_2008_charge_structure}. In this paper, it has been discovered that both gamma-ray reactor feedback and runaway electron transport feedback amplify RREA in the simple reactor, and these feedback mechanisms require a smaller electric field strength to provide self-sustaining RREA development in comparison with the relativistic feedback. In Section \ref{section:simple_reactor} the feedback mechanisms of the simple reactor are described. In Section \ref{section:theory} the reactor feedback is theoretically described. Section \ref{section:simulations} provides Monte Carlo simulations of the simple reactor using GEANT4. In Section \ref{section:discussion} the simple reactor is discussed as a possible mechanism for TGF and TGE, the distinguishing properties of this model that are experimentally observable are considered.

\section{Simple reactor model}
\label{section:simple_reactor}

The reactor feedback is the intensification of RREA development in a thunderstorm supercritical electric field region (cell) by the radiation of other cells \cite{reactor}. The simplest system capable of demonstrating the reactor feedback is the system of two cells with oppositely directed electric fields accelerating electrons toward each other. This system is called ``the simple reactor'' and we consider it as the next step from uniform electric field models to the description of RREA in the electric field of a real thundercloud. In the simple reactor, the distribution of the electric field corresponds to the system of two flat capacitors placed one on top of the other. It can be considered as the approximation of the electric field distribution in the region of the cloud with three charge layers: a positive middle layer and two negative layers. The described electric field distribution can be a part of a natural thunderstorm \cite{Stolzenburg_1994, Marshall_1995, Marshall_1998_estimates, Stolzenburg_1998_structure_2, Stolzenburg_1998_structure_3, Stolzenburg_1998_precipitation, Stolzenburg_2008_charge_structure}.

\begin{figure}
    \centering
    \includegraphics[width=1\linewidth]{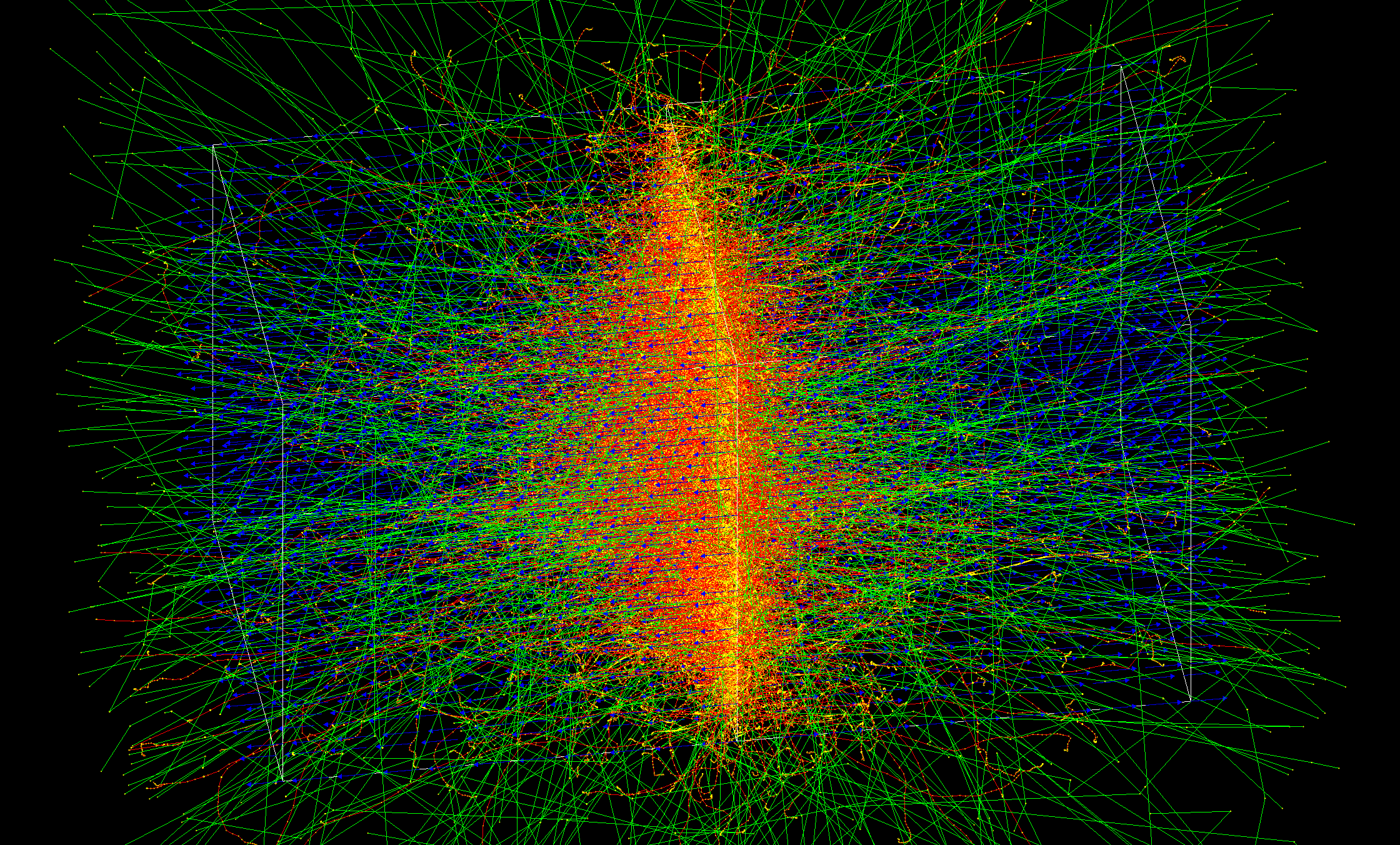}
    \caption{The physics of the simple reactor in the GEANT4 simulation \cite{GEANT4_ALLISON2016}. Green lines - gamma-ray photon tracks, red lines - runaway electron tracks. Blue arrows - electric field lines, yellow dots - particle interaction points. The simulation is started with a single seed electron. The simple reactor consists of two supercritical electric field regions accelerating runaway electrons towards each other. High-energy particles exchange between these regions makes the process self-sustaining. The picture resembles the Eye of Sauron from the Lord of the Rings trilogy: runaway electrons oscillating near the cell boundary form a pupil, the halo of gamma-ray photons and RREAs formed by gamma-ray reactor feedback resembles the cornea of the eye.}
    \label{fig:EYE}
\end{figure}

There are two reactor feedback mechanisms in the simple reactor. The first mechanism, reactor gamma-ray feedback, works in the following way. Let a seed electron form a RREA within one of the cells. RREA grows towards the opposite cell and radiates bremsstrahlung gamma rays \cite{Babich_2020}. Gamma rays have a significant penetration power through the atmosphere at thunderstorm altitudes. Thus, gamma rays reach the opposite cell and propagate through it. Interactions of gamma rays with air molecules of the opposite cell generate seed runaway electrons. These electrons start a new RREA in the opposite cell. Further, the new RREA propagates towards the initial cell, generating gamma rays, which similarly produce RREA in the initial cell. In this way, the process loops, and thus gamma-ray reactor feedback makes RREA self-sustaining in the simple reactor. The second mechanism, runaway electron transport feedback, works in the case when cells are close to each other. In this case, runaway electrons can penetrate the gap between cells, reaching the opposite cell. When a runaway electron reaches the opposite cell, it penetrates inside the cell until the electric field reverses it. After reversal runaway electrons are accelerated toward the initial cell. Thus, in the simple reactor, runaway electrons oscillate near the border of the cells. During the oscillation, runaway electrons are multiplied by the Møller scattering, which also leads to a self-sustaining process. It is established in this paper, that both feedback mechanisms not only can make RREA self-sustaining but also can significantly multiply the number of relativistic particles in the thunderstorm containing the simple reactor (Figure \ref{fig:EYE}). It should be noted that the relativistic feedback naturally impacts RREA development in the simple reactor, however, further it is shown that the influence of the relativistic feedback is negligible compared to the influence of the reactor feedback.

\section{Analytical simple reactor model}
\label{section:theory}
\subsection{Gamma-ray reactor feedback}


To describe the simple reactor gamma-ray feedback theoretically, it is necessary to study the response of a cell to gamma radiation falling into it \cite{reactor}. Let $N$ gamma-ray photons enter the $i$-th cell ($i=$1,2) from above, along the cell's electric field vector. In order to find the feedback coefficient, it should be calculated how much gamma will fly back from the cell towards the other cell. Let $\lambda_{RREA}^i$ be the growth length of an avalanche of runaway electrons, $\lambda_-^i$ be the decay length of gamma radiation, $\lambda_{\gamma}^i$ be the path of a runaway electron before the emission of a gamma-ray photon with supercritical energy, $\lambda_{e^-}^i$ be the path length of a gamma before the birth of a runaway electron, and $P^i$ is the probability of a turn of an electron with further development of the runaway avalanche. These parameters depend on the magnitude of the electric field and the density of the air. In the first approximation, the values for these parameters can be retrieved from the article \cite{STADNICHUK_positron_criterion}. In general, the cells of the simple reactor are assumed to have different field strengths, air density, and cell lengths.

The gamma entering the cell along the electric field will generate electrons with supercritical energy, while losing its energy, which leads to an exponential decay of the primary gamma flux. On the segment [z, z + dz], the flown gammas will give birth to the following number of avalanches of runaway electrons:

\begin{equation}
    \frac{df_{e^-}^i(z)}{dz} dz = N P^i e^{-\frac{z}{\lambda_{-}^i}} \frac{dz}{\lambda_{e^-}^i}
    \label{avalanches_birth}
\end{equation}

The dynamics of the number of bremsstrahlung gamma-ray photons during RREA propagation along the z axis is described by the following equation \cite{STADNICHUK_positron_criterion}:

\begin{equation}
    dN_{\gamma} = e^{\frac{z - z_0}{\lambda_{RREA}}}\frac{dz}{\lambda_{\gamma}} - N_{\gamma} \frac{dz}{\lambda_-}
    \label{eq:dif}
\end{equation}

The first term in \ref{eq:dif} describes the production of bremsstrahlung gamma-ray photons by runaway electrons, and the second term describes a decrease in the number of gamma-rays due to its interaction with air molecules. The solution for \ref{eq:dif} is \ref{eq:sol}:

\begin{equation}
    N_{\gamma}(z, z_0) = \frac{\lambda_{RREA} \lambda_{-}}{\lambda_{\gamma}(\lambda_- + \lambda_{RREA})} \cdot \left(e^{\frac{z - z_0}{\lambda_{RREA}}} - 1\right)
    \label{eq:sol}
\end{equation}

Each RREA grows according to the well-known exponential law \cite{Gurevich1992}, spreading toward the initial gamma rays entering the plane. In this case, depending on the point of birth of the avalanche, the amount of secondary gamma rays that will reach the end of the cell will be as follows (since the avalanche born at the point z will travel a distance equal to z):

\begin{equation}
    \resizebox{0.4\textwidth}{!}{%
        $dF_\gamma^i(z) = \frac{df_{e^-}^i(z)}{dz} dz \frac{\lambda_{RREA} \lambda_{-}}{\lambda_{\gamma}(\lambda_- + \lambda_{RREA})} \cdot \left(e^{\frac{z}{\lambda_{RREA}}} - 1\right)$%
    }
\end{equation}

Thus, the gamma-ray local multiplication factor \cite{reactor} can be calculated (Formula \ref{eq:mult_factor}):

\begin{equation}
    \nu^i = \frac{\int_0^{L^i} \frac{dF_\gamma^i(z)}{dz} dz}{N}
    \label{eq:mult_factor}
\end{equation}

Integration leads to the following formula for the gamma-ray multiplication factor:

\begin{linenomath*}
\begin{equation}
\begin{split}
    \nu^i = \frac{P^i \lambda_{RREA}^i \lambda_-^i}{\lambda_{e^-}^i\lambda_{\gamma}^i(\lambda_{RREA}^i + \lambda_{-}^i)} \\
    \left[ \frac{\lambda_{RREA}^i \lambda_{-}^i}{\lambda_{-}^i - \lambda_{RREA}^i} \left( e^{L^i\frac{\lambda_{-}^i - \lambda_{RREA}^i}{\lambda_{RREA}^i\lambda_{-}^i}} - 1\right) - \lambda_{-}^i \left( 1 - e^{-\frac{L^i}{\lambda_{-}^i}}\right)\right]
\label{local_multiplication_factor}
\end{split}
\end{equation}
\end{linenomath*}

The system with positive feedback can be characterized by the feedback coefficient \cite{Dwyer2007,reactor,STADNICHUK_positron_criterion}. For the simple reactor, the feedback coefficient shows how many times the number of high-energy particles will increase in one full reactor feedback cycle, and it is found with the following formula:

\begin{equation}
    \Gamma = \nu^1 \cdot \nu^2
\end{equation}

The number of particles in the simple reactor grows exponentially with each feedback generation: $N(n) = \Gamma^n$, where n is the number of feedback generation \cite{reactor}. Therefore, the criterion for self-sustaining RREA development in a simple reactor is as follows:

\begin{equation}
    \Gamma = \nu^1 \cdot \nu^2 \geq 1
    \label{criterion_formula}
\end{equation}

\begin{figure}
    \centering
    \includegraphics[width=1\linewidth]{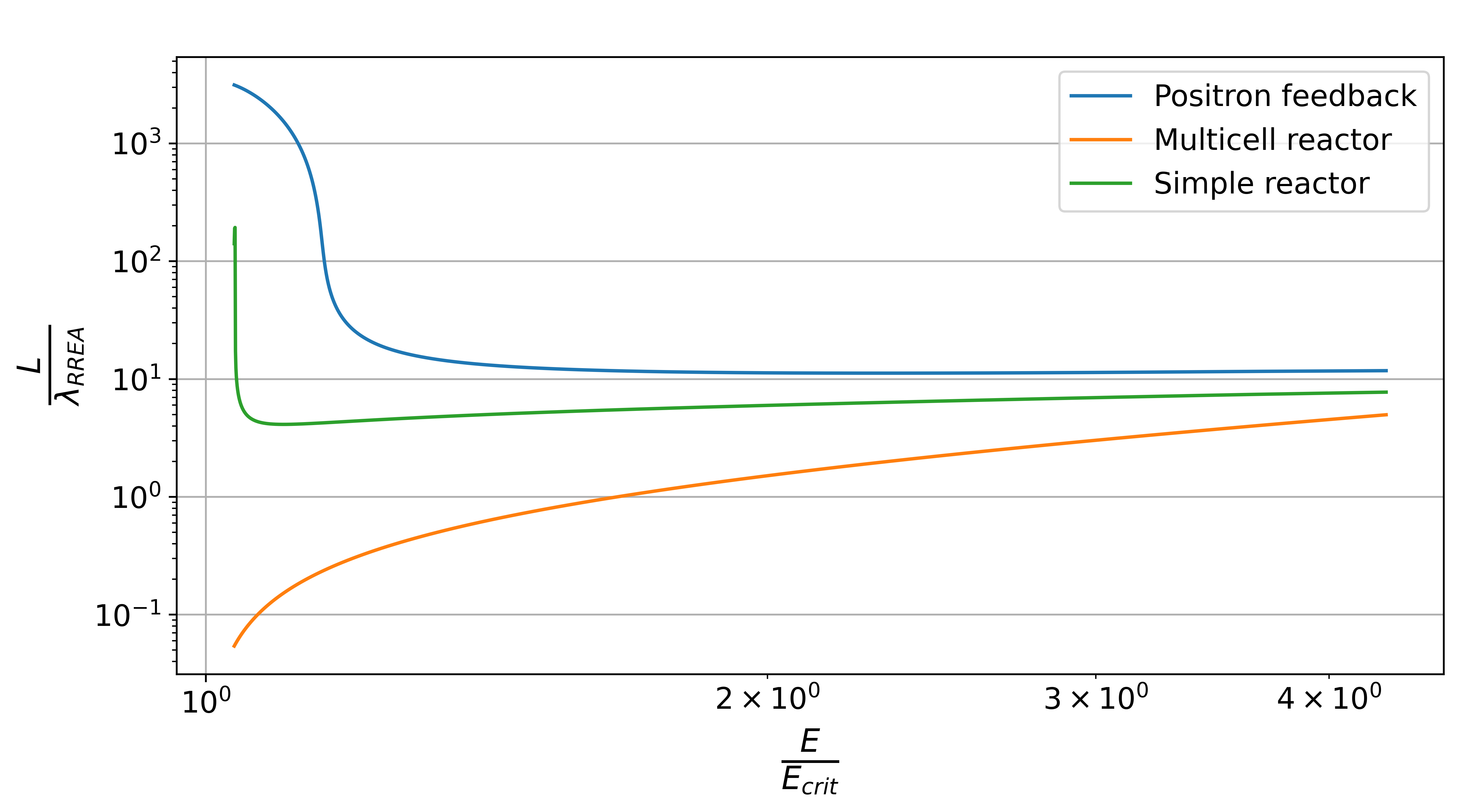}
    \caption{The comparison of self-sustaining positron feedback necessary conditions \cite{STADNICHUK_positron_criterion} and simple reactor self-sustaining feedback necessary conditions (Formula \ref{local_multiplication_factor}, \ref{criterion_formula}). RREA accelerating region length is normalized to $\lambda_{RREA}$, electric field strength is normalized to critical electric field strength (the electric field required for RREA development \cite{Babich_2020}). The simple reactor is also compared with necessary conditions for self-sustaining gamma-ray feedback in multicell reactor model \cite{reactor}. It can be seen that reactor models require significantly lower thunderstorm electric field strengths for self-sustaining RREA development than the relativistic feedback. This important property of the model comes at expense of the complexity of the electric field geometry. The more complex electric field geometry is, the lower electric field strength is required for the feedback to be effective. It should be noted that the conditions for the reactor models are presented without taking into account runaway electron transport between cells. Moreover, it has been discovered that thundercloud hydrometeors can amplify RREA \cite{Zemlianskaya_hydrometeors}. Thus, the exact self-sustaining RREA conditions can be lower than the conditions presented on this picture.}
    \label{simple_reactor_conditions}
\end{figure}

With the obtained criterion, the thunderstorm conditions necessary for self-sustaining RREA development by the reactor gamma-ray feedback can be calculated. These conditions are presented in Figure \ref{simple_reactor_conditions}. Conditions are presented for 3 types of feedback: relativistic positron feedback \cite{Dwyer_2003_fundamental_limit}, simple reactor feedback, and multicell reactor feedback \cite{reactor}. The coordinates in the figure are chosen so that the conditions are invariant with respect to altitude \cite{Dwyer2007,STADNICHUK_positron_criterion}. It can be seen from the figure that both the multicell and the simple reactor feedback mechanisms require significantly lower electric field strength in comparison to the relativistic feedback discharge model. This important property of the reactor feedback comes with a price of the thunderstorm electric field geometry complexity. The most complex electric field geometry, the multicell reactor, requires the lowest electric field strength for the self-sustaining RREA development, while the simplest reactor structure, the simple reactor, requires the electric field strength lying in between the multicell reactor and the uniform electric field. It should be noted that if thundercloud electric field parameters lie above the curve in Figure \ref{simple_reactor_conditions} then the number of energetic particles within the thundercloud grows exponentially \cite{Dwyer2007,reactor,STADNICHUK_positron_criterion}. Otherwise, provided that there is no external source of seed particles, the number of energetic particles decays, and the decay rate depends on the feedback coefficient \cite{reactor}. Thus, even if the feedback does not make the RREA development self-sustaining, it still increases its duration.

\subsection{Runaway electron transport between cells in the simple reactor}

GEANT4 simulations of the simple reactor showed the importance of runaway electron transport between cells for the reactor feedback (Figure \ref{fig:EYE}). In this section, it is shown that the oscillations of runaway electrons between cells in the simple reactor can become self-sustaining and even lead to runaway electron multiplication. At first glance, this effect may seem paradoxical and contrary to the law of energy conservation: While  a single runaway electron move from one cell to another, the total energy it receive from the electric field in the full circle of its oscillation is zero, and, thus, this runaway electron, on average, loses energy in interaction with air. Therefore, a single runaway electron will inevitably lose its energy and stop. However, it is seen in the simulations that runaway electrons oscillate and multiply in the strong electric field of the simple reactor. Therefore, the following question arises: Where do runaway electrons take energy when the feedback becomes self-sustaining? Moreover, the total length of runaway electron motion between cells back and forth cannot be longer than its energy divided by $eE_c$, where $E_c$ - critical electric field, e --- elementary electric charge. This is not more than several tens of meters.

It turns out that the effect of runaway electron transport between cells can be physically explained and that the energy conservation paradox is resolved by runaway electron multiplication. If a runaway electron multiplies by Møller scattering \cite{Dwyer_2003_fundamental_limit}, the result is that the initial and generated runaway electrons receive twice as much energy from the electric field compared to the single initial runaway electron. When the initial runaway electron stops, the secondary electron continues to oscillate and multiply. The reactor feedback in the simple reactor caused by runaway electrons can even become self-sustaining. If the thunderstorm electric field is much stronger than the critical electric field, the runaway electron interaction with the air becomes negligible. Moreover, by the interaction with air, runaway electrons will multiply, which leads to an enormous growth of the number of relativistic particles. Thus, when the electric field strength decreases to values comparable to $E_c$, there is a point where the multiplication of runaway electrons compensates for the energy losses in the air interaction. At this point, the runaway electron transport feedback becomes self-sustaining.

An interesting property of the runaway electron transport feedback is its spatial scale. A runaway electron after hitting an adjacent cell cannot propagate within the cell deeper than its kinetic energy divided by $e(E + E_c)$, where E is the electric field strength of the cell. Therefore, runaway electron transport feedback coefficient depends only on the electric field strength for cell lengths longer than runaway electron maximum energy divided by $e(E + E_c)$, which is about 100 meters for 10 km altitude and 40 MeV maximum energy \cite{Chilingarian_2011_natural_accelerator,asim_spectrum}. Thus, runaway electron transport occurs near the cell interface, which softens the conditions required for self-sustaining RREA development in the simple reactor, because long cells are not needed as in other types of feedback \ref{simple_reactor_conditions}. Nevertheless, it should be noted that runaway electron feedback works effectively only for cells located close to each other since electrons are quickly absorbed by air unless they are accelerated by the electric field.

To theoretically analyze the runaway electron transport feedback, it should first be understood how runaway electrons are decelerated in the electric field of the adjustment cell. Decelerated runaway electron attenuation length can be found by substituting negative electric field strength into the empirical formula for the RREA e-folding length:

\begin{equation}
    \lambda_{decay} = \frac{7300 [kV]}{-E - E_c}
    \label{rrea_decay}
\end{equation}

In this way, number of runaway electrons in the beam will decrease exponentially:

\begin{equation}
    N_{beam}(z) = N_0 e^{\frac{z}{\lambda_{decay}}}
\end{equation}

This analytic continuation of the RREA growth law \cite{Dwyer_2003_fundamental_limit} can be justified in the following way. Normalized runaway electron spectrum can be described with the function:

\begin{equation}
    \frac{df_{RREA}}{d\varepsilon} = \frac{1}{\varepsilon_0} e^{-\frac{\varepsilon}{\varepsilon_0}}
\end{equation}

$\varepsilon_0 = 7.3$ MeV - runaway electron mean energy \cite{Babich_2020}. An electron with energy $\varepsilon$, on average, stops at the coordinate:
 
 \begin{equation}
     z(\varepsilon) = \frac{\varepsilon}{e(E + E_c)}
 \end{equation}
 
 Number of runaway electrons leaving the beam in the interval $(z, z + dz)$ per one primary electron is:

\begin{equation}
    \frac{N_{beam}}{dz} dz = - N_0 \frac{df_{RREA}}{d\varepsilon}\frac{d\varepsilon}{dz} dz = N_0 \frac{E + E_c}{7300[kV]} e^{-\frac{E + E_c}{7300[kV]} z}
\end{equation}

Thus, the formula \ref{rrea_decay} is obtained.

The runaway electron transport feedback coefficient can be defined as the number of runway electrons leaving the cell per one runaway electron entering the cell (analogically to the gamma-ray reactor feedback). The number of runaway electrons, which entered the cell, decreases according to the exponential law derived above as these runaway electrons propagate into the cell (formula \ref{rrea_decay}). When a runaway electron leaves this beam it can stop or it can reverse and form a RREA, which then propagates to the entry plane of the cell. If a RREA starts at the point z, the number of runaway electrons within this RREA reaches $e^{\frac{z}{\lambda_{RREA}}}$ when RREA leaves the cell \cite{STADNICHUK_positron_criterion}. Therefore, if the reversal probability of runaway electrons from the primary beam is equal to P, the runaway electron transport feedback coefficient can be obtained as follows:

\begin{equation}
    \widetilde{\nu}_{e^-} = \int_0^L dz P e^{\frac{z}{\lambda_{RREA}}} \frac{dN_{beam}}{dz} = \frac{P}{\widetilde{\lambda}} \int_0^L dz e^{\frac{1}{\lambda_{RREA}} - \frac{1}{\widetilde{\lambda}}}
\end{equation}

Here $\widetilde{\lambda} = - \lambda_{decay} > 0$. Thus, the following formula is obtained:

\begin{equation}
    \widetilde{\nu}_{e^-} = \frac{P \lambda_{RREA}}{\lambda_{RREA} - \widetilde{\lambda}} \left( 1 - exp\left( L\left(\frac{1}{\lambda_{RREA}} - \frac{1}{\widetilde{\lambda}}\right) \right) \right)
\end{equation}

This formula can be simplified using the empirical formula for $\lambda_{RREA}$ \cite{Dwyer_2003_fundamental_limit}:

\begin{equation}
    \frac{\lambda_{RREA}}{\lambda_{RREA} - \widetilde{\lambda}} = \frac{E + E_c}{2E_c}
\end{equation}

Therefore:

\begin{equation}
    \widetilde{\nu}_{e^-} = P\frac{E + E_c}{2E_c} \left( 1 - exp\left( -L\frac{2E_c}{7300[kV]} \right) \right)
\end{equation}

This formula can be further simplified for cells with cell length $L \gg \frac{7300[kV]}{2E_c}$, which works for cells larger than 100 m:

\begin{equation}
    \widetilde{\nu}_{e^-} = P\frac{E + E_c}{2E_c}
\end{equation}

Generally, there is some space between cells within a thunderstorm. Runaway electrons lose energy by interacting with air molecules while propagating through the gap between cells. A fraction of runaway electrons become undercritical and leave the beam. This fraction can be estimated with the decay length from formula \ref{rrea_decay} for $E = 0$ as $exp\left( -l\frac{E_c}{7300[kV]} \right)$, where $l$ is the gap between cells in the simple reactor. Since, in the first approximation, all runaway electrons lose the same amount of energy in the gap, the shape of their spectrum remains the same. Thus, the formula for runaway electron transport feedback coefficient, taking into account the gap between cells, simply modifies in the following way:

\begin{linenomath*}
\begin{equation}
\begin{split}
    \widetilde{\nu}_{e^-} = P\frac{E + E_c}{2E_c} \left( 1 - \exp\left( -\frac{2E_c L}{7300[kV]} \right) \right) \cdot \\
    \exp\left( -\frac{E_c l}{7300[kV]} \right)
\label{local_multiplication_factor_electrons}
\end{split}
\end{equation}
\end{linenomath*}

\section{GEANT4 simulation}
\label{section:simulations}

The Monte Carlo simulation of the simple reactor was carried out using Geant4, version 4.10.06.p01. Geant4 is recognized as a good tool to model RREA \cite{Skeltved2014,Chilingarian_2018}.
The physics list \texttt{G4EmStandardPhysics\_option4} was chosen as the reliable physics list for RREA simulations \cite{Chilingarian_2018}. This list includes all interactions of electrons, gamma-rays and positrons for energies characteristic for RREA processes \cite{GEANT4_ALLISON2016,reactor}.  The energy cut for the particles was chosen 50~keV based on the fact that low-energy particles will quickly decay, as they do not run away \cite{Babich_2020}, and will not contribute to the feedback. The simulated geometry is a large world volume filled with air, within which a child volume is specified, also filled with air. A simple reactor by definition consists of two child volumes: both volumes are filled with air with a density of 0.414~$kg/m^3$, corresponding to altitude 10~km, and contain electric field in the way that both volumes accelerate runaway electrons towards each other (Figure \ref{fig:EYE}).

The purpose of the GEANT4 simulation is to find the parameters of the system necessary for the self-sustaining RREA regime (when the generation of high-energy particles within the thunderstorm does not stop until the electric field is discharged). The simulation was carried out by varying the cell size and the strength of the electric field inside of it. At a certain electric field strength, the number of gamma-ray photons and runaway electrons crossing in both directions the boundary between the simple reactor cells will not decrease over time. Thus, RREA within the simple reactor will not die out over time. In this case, self-sustaining feedback is reached. Thus, by increasing the electric field with a constant cell length, one can find the critical point at which the reactor will become self-sustaining. In this way, the achievement of critical values is checked, and the conditions are calculated.

\begin{figure}
    \centering
    \includegraphics[width=0.8\linewidth]{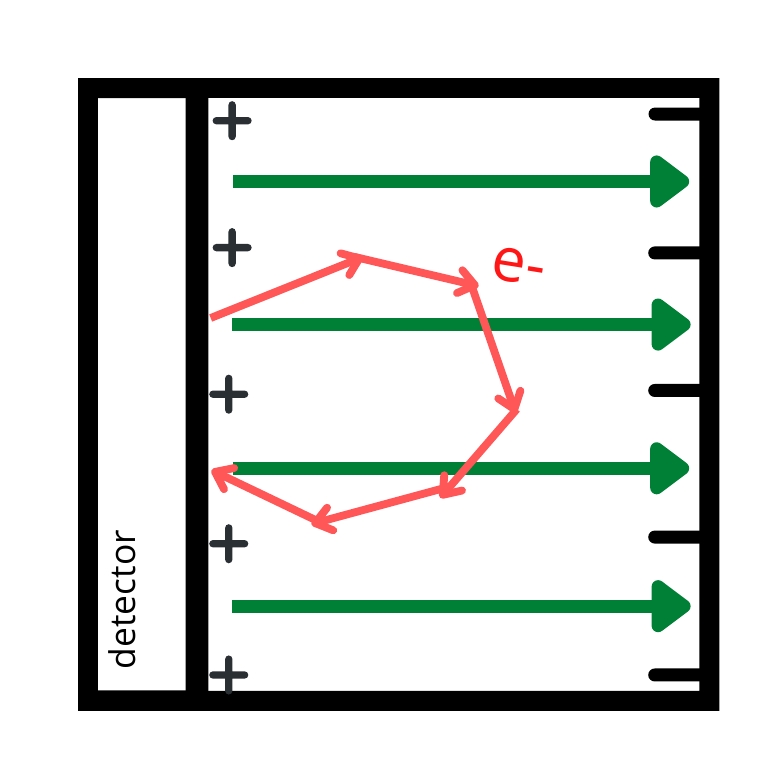}
    \caption{The design of a simple reactor has been simplified in the GEANT4 simulation to consider only one cell as in the figure. Runaway electrons and gamma-ray photons are launched from the right side of the cell along the direction of the electric field. The interactions of launched particles lead to RREA formation, which is accelerated by the electric field to the right side of the cell. In the result, generated particles reach the detector and registered. In the next stages of the simulation, registered particles are launched and new generated particles are similarly registered. In this way, each reactor feedback generation is studied separately, thus, allowing the analysis of the model.}
    \label{model_simple_reactor}
\end{figure}

\begin{figure}
    \centering
    \includegraphics[width=1\linewidth]{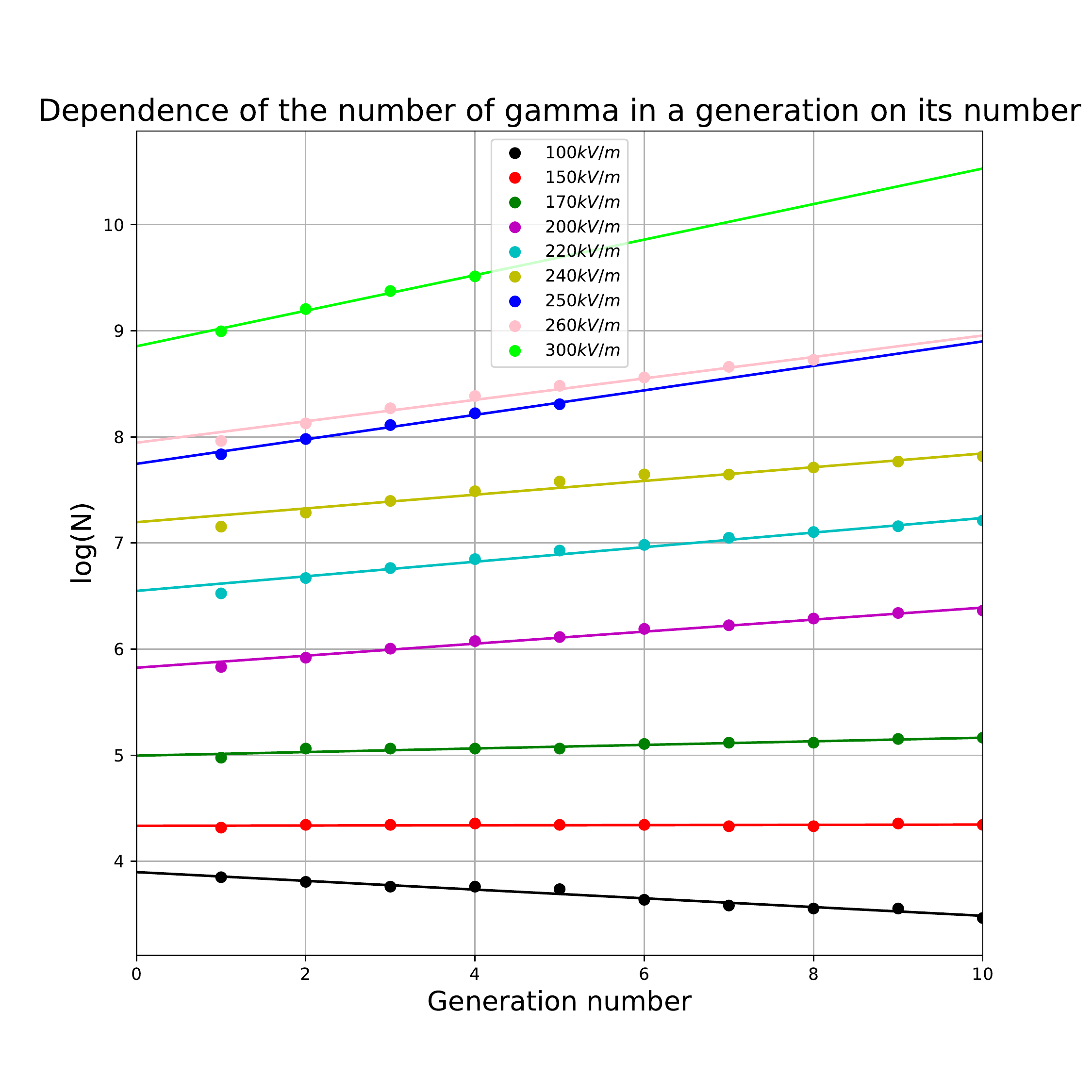}
    \caption{The dependence of the logarithm of the number of gamma-ray photons that propagates from one cell to another in the simple reactor depending on the number of feedback generation. The number of generation is the number of an iteration of a simple reactor simulation (Figure \ref{model_simple_reactor}). It can be seen that the number of gamma-rays produced by the simple reactor exponentially grows or exponentially decays depending on the electric field strength.}
    \label{gamma_generation}
\end{figure}

The most important stage in modeling is the division of the high-energy particles into generations. If directly two cells are created with an oppositely directed field, then it will be quite difficult to divide the process into feedback generations, since, under certain conditions, a self-sustaining feedback is formed and the simulation will not stop. Thus, analogically to the theoretical model, it was decided to simultaneously simulate only a half of a simple reactor. This approach is possible due to the symmetry of the simple reactor. The modeling scheme is shown in Figure \ref{model_simple_reactor}. The model consists of a single cell filled with air and electric field. At the beginning of the cell an air detector is placed --- the volume within which particles are stopped and registered. In the first simulation step, seed particles with an energy of 5~MeV are launched from the beginning of the cell along the direction of the electric field. Seed runaway electrons are decelerated by the electric field. Some of them penetrate into the cell, reverse, and form RREA towards the detector. Seed gamma-ray photons propagate through the cell and interact with air molecules. This interaction results in runaway electrons generation, which reverse and form RREA, also moving and growing toward the beginning of the cell. All particles that reach the detector are stopped and registered and the simulation stops. In this way, the first feedback generation is modeled. In subsequent simulations, the particles registered in the previous iteration are launched into the cell accordingly (thus imitating propagation of the high-energy particles from one cell into another in the simple reactor). These particles interact with the cell, which results in new particles generated that reach the detector. For each feedback generation this process repeats. Figure \ref{gamma_generation} shows the number of gamma-rays reaching the detector in each simulated feedback generation. The graph shows that depending on the thunderstorm conditions the number of high-energy particles can decay from generation to generation or vice versa. The thunderstorm conditions when the number of particles does not change are the necessary conditions for the self-sustaining development of RREA in the simple reactor.

\begin{figure}
    \centering
    \includegraphics[width=1\linewidth]{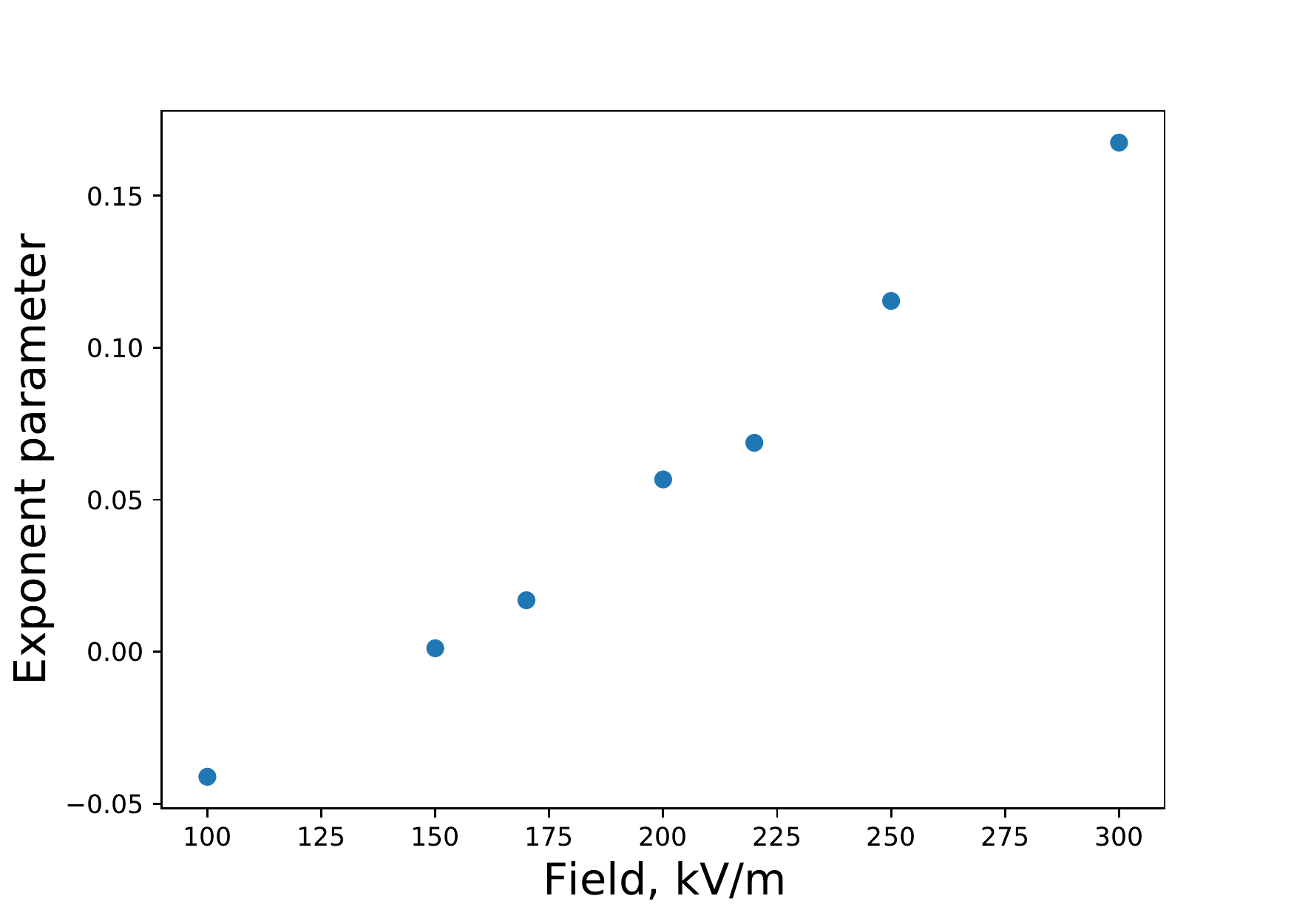}
    \caption{The dependence of the feedback generations exponent parameter on the electric field in the simple reactor for the cell length 400 m. Negative exponent parameters means the decay of RREA in the simple reactor, while positive exponent parameter means self-sustaining RREA development with high energy particles generation. The exponent parameter includes both simple reactor feedback processes: gamma-ray reactor feedback and runaway electron oscillations (Figure \ref{fig:EYE}). The conditions necessary for self-sustainable regime (when exponential parameter equals 0) are in agreement with theoretical predictions (Figure \ref{simple_reactor_conditions}).}
    \label{param_exp_generation}
\end{figure}

To calculate the feedback coefficient, the simulation was launched with seed runaway electrons. Number of generated gamma-ray photons and runaway electrons for each feedback generation was registered, thus forming plots similar to Figure \ref{gamma_generation}. Each plot was fitted with an exponential function. The feedback coefficient \ref{local_multiplication_factor}, \ref{local_multiplication_factor_electrons} is obtained from the coefficient in the exponent by adding 1 to it \cite{reactor}. The obtained dependence of the exponent parameter on the electric field strength is shown in Figure \ref{param_exp_generation}. When the exponent parameter is positive, number of high energy particles in the simple reactor thunderstorm self-sustainably grows until the electric field is discharged.

\begin{figure}
    \centering
    \includegraphics[width=1\linewidth]{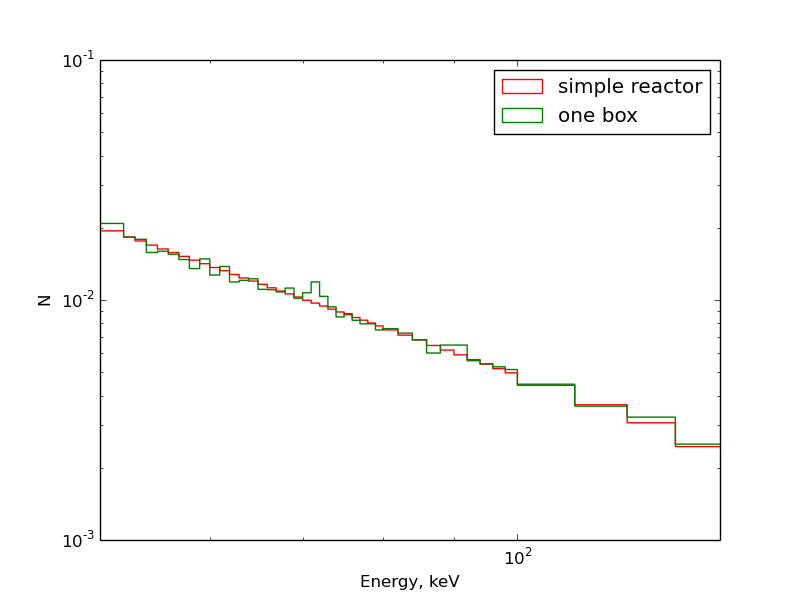}
    \caption{Comparison of the spectra obtained from the simulation of the simple reactor and ordinary RREA spectrum, obtained from a single cell simulation with a uniform electric field. The simulation of the simple reactor was turned off when enough statistics were collected. It can be seen that the spectrum of the simple reactor gamma-radiation is the same as the RREA bremsstrahlung spectrum. It is established that the thunderstorm gamma-radiation spectrum agrees with the RREA spectrum \cite{asim_spectral_analysis,Chilingarian_2020_radon}. Thus, the simple reactor can be one of the mechanisms of TGF and TGE.}
    \label{spectr_compare}
\end{figure}

It is also interesting to calculate the spectrum of the gamma-rays produced within the simple reactor and compare it with the spectrum of an ordinary RREA bremsstrahlung. To obtain the spectrum, a full simple reactor with two cells oriented towards each other was simulated. This simulation captures the particles with their energies in a tracking action. The critical parameters of the simple reactor were chosen --- the field is 300 kV/m and the length of one cell is 400 m for 10 km altitude air density. Similarly to the previous simulation technique, the \texttt{G4EmStandardPhysics\_option4} physics list was used, and the energy cut for particles is 50~keV. The simulation was stopped when the number of high-energy particles reached $10^6$, and the spectrum of registered gamma rays is obtained. In addition, a simulation for a single cell with the same parameters was carried out to obtain the spectrum of an ordinary RREA. The resulting spectra are shown in Figure \ref{spectr_compare}. The graph shows that the spectra are the same. It should be noted that single cell gamma-ray spectrum contains more pronounced positron peak. However, when gamma rays propagate from thunderstorm to the detector registering TGF or TGE, they interact with the atmospheric layer and naturally produce the positron peak. Thus, this peak will also be present when the TGF or TGE produced by the simple reactor is measured. Nowadays it has been reliably established that the TGF and TGE source spectrum is the RREA spectrum \cite{asim_spectrum,Chilingarian_2018}. Thus, the simple reactor can be the mechanism for the TGF or TGE.

\section{Discussion}
\label{section:discussion}

The discovered mechanism called the simple reactor can be applied for a thundercloud containing two regions with electric field exceeding the critical value, i.e. allowing the RREA development (for simplicity, such regions are called cells \cite{reactor}), electric field is oriented in the way that cells accelerate runaway electrons towards each other. It was established that there is a positive feedback in this system caused by two mechanisms (besides the relativistic feedback \cite{Dwyer_2003_fundamental_limit}, which impact is relatively low (Figure \ref{simple_reactor_conditions})). The first mechanism is the transport of runaway electrons from one strong field region to another. This leads to the effective high-energy electron multiplication and runaway electron oscillation near the edge between the strong electric-field regions. The electron transport feedback coefficient is very high for a small gap between cells, and is a dominant RREA multiplication mechanism in the case of the small gap. On the other hand, in the case of a significant gap, when the distance between strong field regions exceeds the characteristic length of runaway electrons, too few runaway electrons propagate through the gap between regions, thus another feedback mechanism dominates. The second feedback mechanism is the gamma-ray reactor feedback \cite{reactor}. RREA bremsstrahlung gamma-rays have high penetration rate in the air. Thus, in the simple reactor, gamma-rays effectively propagate from one cell to another. When a gamma-ray photon propagates through the opposite cell, it interacts with air, producing secondary RREAs, which is the gamma-ray reactor feedback. Both feedback mechanisms can lead to self-sustaining RREA development and, moreover, to rapid multiplication of high-energy particles within a thunderstorm.

The formulas derived in this paper allow one to predict the feedback coefficient for both feedback mechanisms without complicated modeling; the theoretical predictions of this paper are verified by GEANT4. The discovered feedback coefficients completely describe the behavior of the simple reactor, e.g. allow to calculate the conditions required for the self-sustaining RREA development (Figure \ref{simple_reactor_conditions}). The limitations of the proposed analytical model are as follows. Firstly, the model is one-dimensional and, therefore, does not consider the transverse dynamics of the avalanche, which affects the feedback coefficients in the case of narrow electric field regions \cite{STADNICHUK_positron_criterion}. Second, though the description of runaway electron transport feedback qualitatively matches Geant4 simulations, it lacks quantitative accuracy. More theoretical and modeling research is needed to establish the exact influence of the electron transport feedback on the RREA development.

The simple reactor geometry corresponds to the charge distribution with two negative charge layers on both sides of the positive layer. This structure can be a part of a more complicated charge structure of a thunderstorm \cite{Stolzenburg_1994, Marshall_1995, Marshall_1998_estimates, Stolzenburg_1998_structure_2, Stolzenburg_1998_structure_3, Stolzenburg_1998_precipitation, Stolzenburg_2008_charge_structure}. In the simple reactor, the maximum density of runaway electrons will be on the border between two oppositely directed cells  --- in the center of the simple reactor, in the region of the positive charge (it should be noted that the large value of a single positive charge in a region of a cloud can be sufficient for RREA development below and above this region, forming the simple reactor). This feature distinguishes the simple reactor model from models assuming the development of RREA in a single cell with maximum particle density in the cloud top or cloud base. Since in the simple reactor the maximum runaway electron density is located in the center of the reactor, it is harder for bremsstrahlung gamma-rays to reach detectors registering TGF or TGE due to the greater thickness of the atmosphere that they must penetrate. However, this does not contradict the observed gamma-ray fluxes, since the reactor feedback increases the number of generated bremsstrahlung gamma-rays within a thunderstorm containing the simple reactor. This increase compensates for the decrease in gamma-ray flux by extra atmosphere in has to penetrate.

Another distinguishing and important property of the simple reactor is that it generates simultaneous gamma-ray radiation directed upward and downward from a thundercloud (or in other opposite directions if the simple reactor is not oriented vertically). This means that theoretically it is possible to simultaneously detect a TGF or a TGE from two opposite sides of a thunderstorm, e.g., from the top and from the bottom. Such observation can be performed, for example, with an airplane containing particle detectors flying over an observatory with particle detectors. Also a TGF generated by the simple reactor can be registered simultaneously from space and ground observatories, but the probability for the space station to be located above the ground observatory at the moment of TGF is very low due to the TGF short duration. It should be noted that the time profile of the gamma-ray flux in measurements from both sides of the thunderstorm must match in order to conclude that upward and downward gamma-ray radiation are connected by the reactor feedback. This requires a good temporal resolution of the detectors.

The simple reactor with a large feedback coefficient can be a source of TGF. Characteristic timescale of the simple reactor is its size divided by the speed of light, which is in order of microsecond. Therefore, the timescale and radiated gamma-ray spectrum satisfy the experimentally observed TGF data \cite{10_month_ASIM,Fermi_first_results,BATSE_TGF_discovery}. Runaway electron acceleration and its bremsstrahlung gamma-ray radiation in the simple reactor precede the lightning leader and should coincide with the early stage of the lightning initiation. It should be noted that a TGF generated by positive feedback has a characteristic exponential gamma-ray flux rise time profile. Number of high-energy particles grows exponentially on TGF timescales as thunderstorm electric field remains almost constant on these timescales. At the TGF peak, thunderstorm electric field lowers, thus feedback coefficient drops and the feedback becomes finite: the flux of high-energy particles starts to decay or even abruptly terminates, if the electric field required for RREA development abruptly disappear. The disappearance of the electric field can be connected either with local discharges or with the initiation of a lightning leader. From the rise profile of measured TGF flux the feedback coefficient can be restored. The feedback coefficient is a good source of information on the thunderstorm electric field during the TGF (Formula \ref{local_multiplication_factor}, \ref{local_multiplication_factor_electrons}) \cite{STADNICHUK_positron_criterion}.

Another TGF model based on RREA, the relativistic feedback discharge model, supposes significant positive feedback (the relativistic feedback) in the most simple thunderstorm geometry - uniform electric field \cite{Dwyer_2012}. The disadvantage of this model is that it requires very high values of electric field strength extended over a large thunderstorm space \cite{STADNICHUK_positron_criterion,Stadnichuk2019,Babich_2020}. The significant feature of the simple reactor is that it requires  smaller electric field strength for the self-sustaining RREA development than it is in the uniform electric field (Figure \ref{simple_reactor_conditions}) \cite{Dwyer_2012, STADNICHUK_positron_criterion,reactor}. Moreover, provided that two strong field regions are formed by the same positive charge layer, the conditions for self-sustaining feedback in the simple reactor are significantly more achievable than for self-sustaining relativistic feedback.

For the simple reactor (as for any other RREA model with positive feedback \cite{reactor, Dwyer_2003_fundamental_limit}) the following time dependence of the gamma radiation flux measured on the ground is possible. Usually during a TGE measurement, the gamma flux slowly increases exponentially \cite{Chilingarian_2020_radon, Wada2019}. This can be explained by the fact that when the cloud approaches the detector at a constant speed, so the distance from the cloud to the TGE source decreases linearly in time. The measured particle flux decays exponentially with distance, thus, if the distance is decreased linearly, the measured flux grows exponentially \cite{Wada2019}. If RREAs are self-sustaining within the thunderstorm due to the positive feedback, then their bremsstrahlung gamma-ray flux grows exponentially within the thunderstorm itself (it can grow slowly if the multiplication rate is slightly higher than unity). Moreover, even if the feedback is present but the RREA is not self-sustaining due to the low feedback coefficient, the RREA time profile is modified and its radiation time increases \cite{reactor}. Thus, with the positive feedback, the time profile of the measured gamma-ray flux is exponent superimposed on exponent. The time profile can be more complicated if the electric field within thunderstorm is changing. Such time profile was measured during winter thunderstorms gamma-ray glows \cite{Wada2019}, which supports the hypothesis about the importance of the positive feedback in thunderstorm physics.

Lightning initiation by RREA is a widely discussed problem in the atmospheric electricity science community \cite{kostinsky_2020_mechanism_2, kostinsky_2019_mechanism_1,Gurevich_2001,Gurevich1992,Dwyer_2012,Chilingarian_2020_radon,reactor,Mezentsev_2022_TGF_lightning}. Within the simple reactor, RREAs are directed to the center of the system, thus creating the maximal density of RREA electrons and their products in the center. This also leads to maximum ionization in the middle part of the simple reactor \cite{Khamiton2020,Dwyer_Babich_2011}. The described case can be more favorable for streamer initiation when compared to a single strong field region with RREAs directed to the top or to the bottom base of a cloud because the ionization has its maximum at the end of a RREA, on the edge of the strong field region. Moreover, the simple reactor naturally contains more high-energy particles than the uniform electric-field region because of the reactor feedback. Thus, the simple reactor model can be a useful mechanism for lightning initiation research. It should be noted, that if streamers are generated with the reactor feedback, it can lead to an exponential growth of radio signal preceding the lightning leader.

\section{Conclusion}

This paper studies RREA physics in thunderstorms containing two supercritical electric field regions accelerating runaway electrons toward each other. Such a system, named the simple reactor, can be a part of a natural thunderstorm. It is discovered that RREA in the simple reactor has positive reactor feedback. The reactor feedback enhances RREA duration and can lead to self-sustaining RREA development. There are two mechanisms of the reactor feedback in the simple reactor. RREA is effectively multiplied by the gamma-ray exchange between regions even if they are far enough apart. If regions are close to each other, high-energy particles are generated by the runaway electron oscillations near the border between regions. In this case, the small-scale strong electric field is sufficient for self-sustaining RREA development. It is shown that the reactor feedback in the simple reactor requires significantly lower electric field strength for RREA multiplication compared to relativistic feedback.

The simple reactor in the self-sustaining regime rapidly increases the number of high-energy particles within a thunderstorm and can hypothetically precede or cause lightning initiation. It is established that the time scale and the spectrum of the simple reactor gamma radiation agree with TGF data. The distinguishing property of the simple reactor is that it radiates gamma rays in two opposite directions. This allows simultaneous and correlated observation of TGF or TGE gamma rays from the top and from the bottom of a thundercloud. Moreover, the feedback coefficient can be retrieved from TGF and TGE data, which can be a good source of information about gamma radiating thunderstorm parameters, including electric field strength, supercritical region length, and the electric field geometry.

\section*{Acknowledgements}

The work of E.~Stadnichuk was supported by the Foundation for the Advancement of Theoretical Physics and Mathematics “BASIS”. The work of E.~Svechnikova was supported by a grant from the Government of the Russian Federation (contract no. 075-15-2019-1892).

\bibliography{apssamp}

\end{document}